# EVOLUTION OF THE SCIENCE FICTION WRITER'S CAPACITY TO IMAGINE THE FUTURE

Liane Gabora
University of British Columbia
Department of Psychology, Okanagan Campus
Fipke Centre for Innovative Research, 3247 University Way
Kelowna BC, Canada V1V 1V7
Email: liane.gabora@ubc.ca

**ABSTRACT**

Drawing upon a body of research on the evolution of creativity, this paper proposes a theory of how, when, and why the forward-thinking story-telling abilities of humans evolved, culminating in the visionary abilities of science fiction writers. The ability to recursively chain thoughts together evolved approximately two million years ago. Language abilities, and the ability to shift between different modes of thought, evolved approximately 100,000 years ago. Science fiction dates to at least the second Century AD. It is suggested that well before this time, but after 100,000 years ago, and concurrent with the evolution of a division of labour between creators and imitators there arose a division of labour between past, present, and future thinkers. Agent-based model research suggests there are social benefits to the evolution of individual differences in creativity such that there is a balance between novelty-generating creators and continuity-perpetuating imitators. A balance between individuals focused on the past, present, and future would be expected to yield similar adaptive benefits.

**INTRODUCTION**

Science fiction writers possess an often uncanny ability to envision the future before it happens, particularly with respect to technological advances. Drawing upon research in psychology, anthropology, archaeology, and agent-based modeling, this paper offers a theory of how these abilities evolved. First, we look at two key cognitive transitions that have been proposed to underlie the uniquely creative abilities of humans. Next, we examine evidence that individual differences in creativity are adaptive at the level of the social group. Third, I argue that, using analogous reasoning, we could expect individual differences in the proclivity to focus one's thoughts along the spectrum from past, to present, to future, should also be adaptive at the level of the social group.

**THE EVOLUTION OF HUMAN CREATIVITY**

We now outline a body of research on the evolution of creativity that will form the scaffold for the rest of this paper. We will examine how, when, and why the forward-thinking story-telling abilities of humans evolved, culminating in the visionary abilities of science fiction writers.

**Recursive Recall and the Chaining of Thoughts**

How did the capacity for human creativity evolve in the first place? Let us first consider how the mind acquired the capacity to modify thoughts and ideas by thinking about them in the context of other thoughts and ideas that are similar, that is, in the same local cluster. Merlin Donald (1991) suggested that the enlarged cranial capacity of our *Homo erectus* ancestors 1.7 million years ago enabled them to voluntarily retrieve and modify memories independent of environmental cues (sometimes referred to as 'autocuing'), a capacity he referred to as *self-triggered recall and rehearsal,* and which ushered forth a transition to a new mode of cognitive functioning. Thus, while *Homo habilis* was limited to the "here and now", *Homo erectus* could *chain* memories, thoughts, and actions into more complex ones, and progressively modify them, thereby gaining new perspectives on past or possible events, and even mime or re-enact them for others. The notion of self-triggered recall bears some resemblance to Hauser et al.'s (2002) idea that what distinguishes human cognition from that of other species is the capacity for recursion, and to Penn, Holyoak, and Povinelli's (2008) concept of *relational reinterpretation*, the ability to reinterpret higher order relations between perceptual relations.

Donald's proposal has been shown to be consistent with the structure and dynamics of associative memory (Gabora 2000, 2010, 2017). Neurons are sensitive to primitive stimulus attributes or 'microfeatures', such as sounds of a particular pitch or lines of a particular orientation. Experiences encoded in memory are *distributed* across cell assemblies of neurons, and each neuron participates in the encoding of many experiences. Memory is also *content-addressable:* similar stimuli activate overlapping distributions of neurons. With larger brains, experiences could be encoded in more detail, enabling a transition from coarse-grained to fine-grained memory. Fine-grained memory enabled concepts and ideas to be encoded in more detail, that is, there were more ways in which distributed sets of microfeatures could overlap. Greater overlap enabled more routes for self-triggered recall, and paved the way for streams of abstract thought. Ideas could now be reprocessed until they fit together with cognitive structures already in place, allowing for the emergence of local clusters of mutually consistent ideas, and thus for a more coherent internal model of the world, or worldview (Gabora 1999; Gabora and Aerts 2009; Gabora and Steel 2017). This in turn paved the way for a primitive form of storytelling, although it was limited to mime and gesture, as complex language had not yet evolved.

In short, it is suggested that the onset of creative cultural evolution, including a capacity for simple mime and gesture based storytelling, was made possible by the onset of the capacity for one thought to trigger another, leading to the chaining and progressive modification of thoughts and actions. However, due the sparseness of the pre-modern archaeological record, it is difficult to experimentally test hypotheses about how the creative abilities underlying cultural transitions evolved. Although methods for analyzing archaeological remains are becoming increasingly sophisticated, they cannot always distinguish amongst different theories.

Agent-based modeling is a computational methodology in which artificial agents can be used to represent interacting individuals. It enables us to address questions about the workings of collectives such as societies. It is particularly valuable for answering questions of this sort which lie at the interface between anthropology and psychology, owing to the difficulty of experimentally manipulating a variable, such as the average amount by which one invention differs from its predecessor and observing its impact on cumulative culture over time.

EVOC (for EVOlution of Culture) is a computational modeling of cultural evolution that consists of neural network based agents that invent new actions and imitate actions performed by neighbors (Gabora 1995, 2008b). The assemblage of ideas changes over time not because some replicate at the expense of others, as in natural selection, but through inventive and social processes. Agents can learn generalizations concerning what kinds of actions are useful, or have a high 'fitness', with respect to a particular goal, and use this acquired knowledge to modify ideas for actions before transmitting them to other agents. A model such as EVOC is a vast simplification, and results obtained with it may or may not have direct bearing on complex human societies, but it allows us to vary one parameter while holding others constant and thereby test hypotheses that could otherwise not be tested. It provides new ways of thinking about and understand what is going on.

The hypothesis that cultural evolution was made possible by the onset of the capacity for one thought to trigger another was tested in EVOC by comparing runs in which agents were limited to single-step actions to runs in which they could chain ideas together to generate multi-step actions (Gabora, Chia, and Firouzi, 2013; Gabora and Smith, submitted). Chaining increased the mean fitness and diversity of cultural outputs across the artificial society (Gabora, Chia, and Firouzi, 2013). While chaining and no-chaining runs both converged on optimal actions, without chaining this set was static, but with chaining it was in constant flux as ever-fitter actions were found. While without chaining there was a ceiling on mean fitness of actions, with chaining there was no such ceiling, and chaining also enhanced the effectiveness of the ability to learn trends. These findings supported the hypothesis that the ability to chain ideas together can transform a culturally static society into one characterized by open-ended novelty.

**Contextual Focus and Language**

To recap so far: it is suggested the evolution of the capacity for science fiction, and storytelling more generally, had its roots 1.7 million years ago in the onset of the capacity to chain thoughts and actions together and thereby string events into narratives. However, the only means of sharing such narratives with others was to express them through mime, i.e., act them out. Thus, the earliest forms of storytelling are thought to be oral, in conjunction with gestures and expressions (Banks-Wallace 2002). It was possible to think about an idea in relation to other closely related ideas and thereby forge clusters of mutually consistent ideas, which allowed for a narrow kind of creativity, limited to minor adaptations of existing ideas. However, the mind was not integrated, nor truly creative, until it could forge connections between seemingly disparate ideas as in the formation of analogies.

The Middle-Upper Paleolithic has been referred to as the birth of art, science, and religion, and the 'big bang' of human creativity (Mithen 1998). Although the timing, location, and abruptness of this shift has been the subject of extensive debate (e.g., McBrearty and Brooks 2000), it is evident that something took place around this time. The question is: what caused it?

One proposal is that it was due to the onset in the Middle/Upper Paleolithic of *contextual focus* (CF): the ability to shift between different modes of thought—an explicit *analytic mode* conducive to logical problem solving, and an implicit *associative mode* conducive to insight and breaking out of a rut (Gabora 2003). While dual processing theories generally attribute abstract, hypothetical thinking solely to the more recently evolved "deliberate" mode (e.g., Evans, 2003), according to the CF hypothesis it is possible in either mode but it will differ character in the two modes (flights of fancy versus logically constructed arguments) (Sowden, Pringle, and Gabora 2014). CF thus paved the way for integration of different domains of knowledge (Mithen 1998).

It has been proposed that CF was made possible by mutation of the FOXP2 gene, which is known to have undergone human-specific mutations in the Paleolithic era (Chrusch and Gabora 2014; Gabora & Smith, submitted). FOXP2, once thought to be the "language gene", is not uniquely associated with language. The idea is that, in its modern form, FOXP2 enabled fine-tuning of the neurological mechanisms underlying the capacity to shift between processing modes by varying the size of the activated region of memory.

The hypothesis that the onset of CF brought about a second cognitive transition underlying the human capacity to evolve complex culture was also tested with EVOC (Gabora, Chia, and Firouzi 2013; Gabora and Smith submitted). When the fitness of an agent's outputs was low it temporarily shifted to a more divergent mode by increasing $\alpha$: the degree to which a newly invented idea deviates from the idea on which it was based. Both mean fitness of actions across the society increased with CF, as hypothesized, and CF was particularly effective when the fitness function changed, which supported its hypothesized utility in breaking out of a rut and adapting to new or changing environments. Using an entirely different computational architecture, CF was similarly shown to enhance the art-making abilities of a computational creativity program geared at generating portraits with painterly qualities (DiPaola and Gabora, 2009; Gabora and DiPaola, 2012).

The evolution of the capacity for CF enabled or ancestors to control their thought processes—effectively tailor them to the task at hand—examining their inner and outer worlds from not just different perspectives but at different hierarchical levels (e.g., from detailed to 'big picture'). This enabled them to connect seemingly unrelated aspects of their lives into a more integrated understanding of their world, and it enabled the evolution of complex language. Thus, it made it possible to go from expressing stories by acting them out to *telling* stories. In addition to being part of religious rituals, some archaeologists believe rock art, and tattooing may have served as a form of storytelling in ancient cultures (Kaeppler 1988; Lewis-Williams et al. 1982).

Storytelling is something that, to some degree, we are all capable of; however, not all of us are equally interested in, nor good at, telling stories. I suggest that some other hurdles had to be crossed in the transition to a storytelling species, and in particular, a species that generates science fiction.

**INDIVIDUAL DIFFERENCES IN CREATIVITY**

Although creativity is encouraged in the abstract it is often discouraged in educational and workplace settings, suggesting that there may be corrective forces at work in society that temper the novelty-generating effects of creativity with the continuity-promoting effects of imitation and ritual (Gabora and Tseng 2017). Such corrective forces might be expected to exert a stronger impact on those who show less creative potential, thereby giving rise to a different degrees and kinds of creativity. Indeed, there *are* pronounced individual differences in creativity, not just in terms of domain of application but also in terms of degree and scope (Chen, Himsel, Kasof, Greenberger and Dmitrieva 2006. Wolfradt and Pretz 2001; Woodman and Schoenfeldt 1989).

Using the above-mentioned agent-based model of cultural evolution (EVOC), we investigated the idea that tempering the novelty-generating effects of creativity with the novelty-preserving effects of imitation is beneficial for society (Gabora and Tseng 2014a,b, 2017; Leijnen & Gabora 2009). Although the model is vastly simpler than real societies it enabled us to manipulate the ratio of creators to imitators and the degree to which creators are creative in a controlled manner and observe the result.

In a first experiment, we systematically introduced individual differences in creativity, and observed a trade-off between the ratio of creators to imitators and how creative the creators were. Because a proportion of individuals benefit from creativity without being creative themselves by imitating creators, the rate of cultural evolution increases when the novelty-generating effects of creativity are tempered with the novelty-preserving effects of imitation. If there were few creators they could afford to be more creative, and vice versa; if there were many their creativity had to be restrained to exert the same global benefit for the society. Excess creativity was detrimental because creators invested in unproven ideas at the expense of propagating proven ones.

We also obtained evidence that society can benefit by rewarding and punishing creativity on the basis of creative success. In a second experiment, we tested the hypothesis that society as a whole benefits if individuals adjust how creative they are in accordance with the fitness of their creative outputs. I refer to this as *social regulation* because could be mediated by social cues such as praise and/or criticism from peers, family, or teachers, but it is also possible that it involves individual differences in the ability to detect or respond to such cues, or individuals' own assessments of the worth of their ideas, or some combination of these. In the *social regulation* condition of our simulation, each agent regulated its invention-to-imitation ratio as a function of the fitness of its cultural outputs; thus, effective creators created more, and ineffective creators created less. With social regulation, the agents segregated into creators and imitators, and the mean fitness of outputs was temporarily higher. We hypothesized that the temporary nature of the effect was attributable to a ceiling on output fitness.

This in turn led to the hypothesis explored in a third experiment, which explored the conditions under which the benefits of social regulation of creativity are long-term. In keeping with the research discussed earlier suggesting that onset of the capacity for chaining was a pivotal transition in the evolution of human creativity, this third experiment made the space of possible outputs open-ended by giving agents the capacity to chain simple outputs into arbitrarily complex ones. This meant that fitter outputs were always possible, and thus the space of possibilities was in theory unlimited. With social regulation *and* the capacity for chained outputs, the agents once again segregated into creators and imitators, and the mean fitness of their outputs was higher. However, as hypothesized, the effect of social regulation was no longer temporary; it could indeed be sustained indefinitely. We did not test the effect of adding the capacity for contextual focus in this particular set of experiments, but our previous results suggest that it would have magnified the effect of social regulation to increase the mean fitness of cultural outputs further still.

Together, these experiments provide evidence that individual differences in creativity are of not just temporary but ongoing adaptive benefit to society, that these benefits can be that they could come about and be maintained due to social regulation mechanisms. Although further investigation is needed to establish the relevance of these results to real societies, they are a step forward to understanding the underlying mechanisms that enable societies to balance novelty with continuity.

It is difficult to pinpoint when the capacity for a division of labour between creators and imitators across societies could have arisen. What we can say is that it was after the 'big bang of human creativity in the Paleolithic. In any case, with the arrival of the internet, individual differences in the expression of creativity has exploded, with respect to both degree and domain. For any particular topic, every community seems to have *someone* who is an expert on it, and conversely, everyone seems to be an expert on *something*.

**PAST, PRESENT, AND FUTURE THINKING AS A FORM OF INDIVIDUAL DIFFERENCES**

We have examined evidence that individual differences in the balance between novelty-generating creators and continuity-perpetuating imitators may have adaptive benefits for society at large. This suggests that there may have group selection pressure to preserve and perhaps amplify individual differences in creativity over time. Let us take now this line of reasoning one step further.

A balance between individuals focused on the past, present, and future would be expected to yield similar adaptive benefits. Those who are focused on the here and now would be more apt to detect the presence of predators or food items, signs of illness or weather changes, and so forth. Those who are focused on the past would be better able to provide a stabilizing sense of continuity, and to make use of past lessons to avoid repeating mistakes. Finally, those who are focused on the future would help society prepare for effects to come, and think in terms of not just short-term benefits but long-term goals. Ostensibly, a society that consisted of individuals along the spectrum from past to future might argue more, because their points of view will not always be in sync. But nevertheless it is easy to see why such a society would be more successful.

Thus, it is suggested that well before the earliest known works that could be called science fiction in second Century AD, but after 100,000 years ago, and concurrent with the evolution of a division of labour between creators and imitators discussed above, there arose a division of labour between past, present, and future thinkers, as illustrated schematically in Figure 1. The eye in the middle represents someone who naturally focuses on, and thinks most clearly about, the present. The eye on the left represents someone who naturally focuses on, and thinks most clearly about, the past. The eye on the right represents someone who naturally focuses on, and thinks most clearly about, the future.

Most people are probably of the sort that they focus on, and think most clearly about, the present. They are *capable* of thinking about the past and the future, but this is not the natural comfort zone or 'attractor state' for their thoughts; for them both the past and the future are much hazier than what is going on now.

Archaeologists, historians, writers of historical fiction, and so forth, are more likely to focus on, and think most clearly about, the past. They tend to view the present and future in terms of how it is rooted in what has come before.

Futurists, inventors, and writers of science fiction, are more likely to focus on, and think most clearly about, the future. It is not that they cannot or do not think about the past or present but, that they tend to view the past and present as seeds for speculation and prediction about what has yet to pass. Thus, it is proposed that the evolution of individual differences in the extent to which we focus along the spectrum from past to present to future paved the way for the fantastical stories of future events and far-off worlds that we now enjoy.

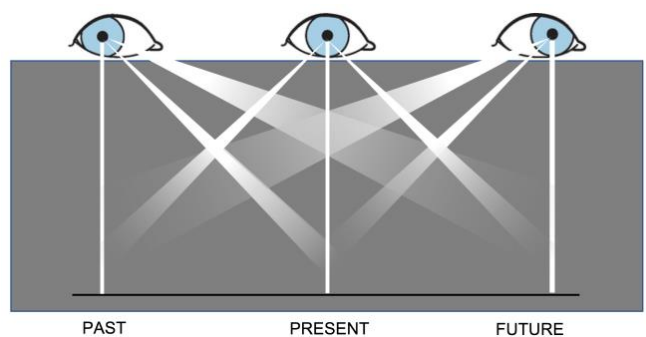

Figure 1. Schematic Depiction of Individual Differences in Tendency to Dwell on Past, Present, and Future

As with the evolution of individual differences in creativity, it is difficult to pinpoint when the capacity for a division of labour between creators and imitators across societies could have arisen. What we can say is that it was after the 'big bang of human creativity in the but before the earliest known works that could be called science fiction in the second Century AD.

**THE BIRTH OF SCIENCE FICTION**

Science fiction, which dates to at least the second Century AD, is a genre of speculative fiction that typically deals with imaginative concepts such as futuristic science and technology, parallel universes, extraterrestrial life, and travel through time and space, sometimes at faster than light speeds. As such, more than other forms of storytelling, it would attract and rely upon individuals whose thought processes lean toward thinking about the future.

**CONCLUSION**

The 'divide and conquer' strategy is well-known to Mother Nature, and it has previously been suggested that its effectiveness can account for individual differences in human creativity. Using a similar argument, this paper suggested that individual differences in the tendency to focus one's thoughts on the past, present, or future, became magnified over time. This in turn paved the way for forward-thinking science fiction writers and their often uncanny powers to envision technological advances before they become reality.

**FUTURE RESEARCH**

This account, though built on an extensive foundation of research in psychology, anthropology, archaeology, and agent-based modeling, is at this point speculative. In future agent-based model work we will investigate the extent to which a division of labour into past, present, and future focused modes of cognition exists and is in fact beneficial to a social group. If so, this would provide tentative support for the hypothesis that individual differences in the tendency to focus one's thoughts on the past, present, or future, became magnified over time yielding benefits for societies. This in turn would provide further support for the hypothesis that such differences led to the rich treasure trove of science fiction that has inspired us for generations and will continue to inspire us for generations to come.

**ACKNOWLEDGEMENTS**

The author acknowledges funding from grant (62R06523) from the Natural Sciences and Engineering Research Council of Canada.

## BIOGRAPHY

**LIANE GABORA** s a Professor in the Department of Psychology at the Okanagan Campus of the University of British Columbia. Her research focuses on the mechanisms underlying creativity, and how creative ideas—and culture more generally—evolve, using a combination of computational modeling and empirical studies with human participants. She has almost 200 articles published in scholarly books, journals, and conference proceedings, has procured over one million dollars in research funding, supervised numerous graduate and undergraduate students, and given talks worldwide on creativity and related topics. She has a short story titled 'Violation' published in *Fiction*, and another titled 'One Way Trip' forthcoming in *Fiddlehead*. She is working on a novel titled *Quilandria* that merges her scholarly and creative writing interests. She studied creative writing at Humber College, Toronto, and the University of California, Berkeley.

Homepage: https://people.ok.ubc.ca/lgabora/